\documentclass[]{aa}
\usepackage{graphicx}
\begin{document}

\title{{\it Research Note:} New Variable Stars in Open Clusters I: Methods and Results
for 20 Open Clusters\thanks{Based on 
observations obtained at Complejo Astron\'omico el Leoncito (CASLEO), 
operated under the agreement between the
Consejo Nacional de Investigaciones Cient\'ificas y T\'ecnicas de la Rep\'ublica 
Argentina and the National Universities of La Plata, C\'ordoba y San Juan;
ESO-La Silla and UTSO-Las Campanas}}
\author{E.~Paunzen\inst{1,2}, K.~Zwintz\inst{1}, H.M.~Maitzen\inst{1},
O.I.~Pintado\inst{3}, M.~Rode-Paunzen\inst{1}}

\mail{Ernst.Paunzen@univie.ac.at}

\institute{Institut f\"ur Astronomie der Universit\"at Wien,
           T\"urkenschanzstr. 17, A-1180 Wien, Austria
\and       Zentraler Informatikdienst der Universit\"at Wien,
           Universit\"atsstr. 7, A-1010 Wien, Austria
\and	   Departamento de F\'isica, Facultad de Ciencias Exactas 
           y Tecnolog\'ia, Universidad Nacional de Tucum\'an, Argentina - Consejo Nacional 
		   de Investigaciones Cient\'ificas y T\'ecnicas de la Rep\'ublica Argentina}

\date{Received 2003/ Accepted 2004}
\authorrunning{Paunzen et al.}{}
\titlerunning{New Variable Stars in Open Clusters I.}{}
\abstract{We present high precision CCD photometry of 1791 objects in 20 open clusters
with an age from 10\,Myr to 1\,Gyr.
These observations were performed within the $\Delta a$ photometric system which is
primarily used to detect chemically peculiar stars of the upper main sequence.
Time bases range between 30 minutes and up to 60 days with 
data from several nights. We describe the way of time series analysis reaching a detection
limit of down to 0.006\,mag.
In total, we have detected 35 variable objects from which four are not members of their
corresponding clusters. The variables cover the entire Hertzsprung-Russell-diagram, hence 
they are interesting targets for follow-up observations.
\keywords{Open clusters and associations: general -- stars: variables}
}
\maketitle

\section{Introduction}

The detection of variable members in open clusters is very important since
these objects have fairly well known astrophysical parameters, such as 
luminosity and effective temperature. Several theories (e.g. pulsational
and evolutionary models) can be tested with these variable stars. 

In the literature, a huge amount of papers dedicated to the search for new
variable stars in open clusters can be found. In general, two different kinds of surveys
are conducted: 1) the search for special types of variables 
(Viskum et al. 1997, Jerzykiewicz et al. 2003) or 2) selected open clusters
are searched for all kinds of variable objects (Kafka \& Honeycutt 2003, 
Mochejska et al. 2003). 

Our search for new variable stars in open clusters is a serendipity result
from already published CCD $\Delta a$ photometry (Bayer et al. 2000, 
Paunzen \& Maitzen 2001, 2002 and Paunzen et al. 2002, 2003).
The intermediate band, three filter $\Delta a$ system investigates the flux
depression at 5200\,\AA\, found for magnetic chemically peculiar
objects (Maitzen 1976). Our observations span widely different time intervals (0.02 to 60 days)
yielding different possibilities for detecting
the whole set of variations. We want to emphasize
that these observations are not optimized for the detection of variable stars
but are capable to find even very low amplitude variables (the typical detection
limit reached is between 0.006 and 0.022 mag).  

We describe the way to define the variability limit and present all bona-fide
variable stars within the Hertzsprung-Russell-diagram. Four objects are
probably not members of the corresponding open clusters. We give a discussion about the
possible nature of the detected variability.

\begin{table*}
\caption[]{Open clusters observed at ESO and UTSO in 1995 (upper panel)
as well as CASLEO in 1998 and 2001 (lower panel). The ages (log\,$t$) and 
distance moduli ($V_0-M_V$) were taken from the literature. The limit
of apparent variability (Limit) is according to Sect. \ref{time}.
The errors in the final digits of the corresponding quantity
are given in parentheses.}
\label{clusters}
\begin{center}
\begin{tabular}{lcrccccrccr}
\hline
\hline
\multicolumn{2}{c}{Designation} & N$_{S}$ & N$_{V}$ & N$_{F}$ & N$_{N}$ & 
JD (start) & \multicolumn{1}{c}{$\Delta t$} & Limit & log\,$t$ & $V_0-M_V$ \\
& & & & & & & \multicolumn{1}{c}{[d]} & [mag] \\
\hline
NGC 2439 & C0738$-$315 & 115 & 3 & 18 & 4 & 2449816.51736 & 7.995 & 0.022 & 7.30 & 13.00(10) \\
NGC 2489 & C0754$-$299 & 53 & 1 & 13 & 4 & 2449818.57083 & 34.937 & 0.012 & 8.45 & 10.80(10)  \\
NGC 2567 & C0816$-$304 & 34 & $-$ & 17 & 4 & 2449818.61111 & 44.868 & 0.012 & 8.43 & 11.10(10) \\
NGC 2658 & C0841$-$324 & 84 & 1 & 12 & 3 & 2449817.56597 & 18.026 & 0.018 & 8.50 & 12.90(15) \\
Melotte 105 & C1117$-$632 & 122 & $-$ & 15 & 2 & 2449816.61111 & 3.072 & 0.010 & 7.77 & 11.30(10) 
\\
NGC 3960 & C1148$-$554 & 32 & $-$ & 17 & 3 & 2449828.60625 & 59.865 & 0.014 & 8.88 & 11.10(20) \\
NGC 5281 & C1343$-$626 & 16 & 1 & 55 & 6 & 2449816.78125 & 48.010 & 0.008 & 7.04 & 10.60(15) \\
NGC 6134 & C1624$-$490 & 82 & 2 & 32 & 6 & 2449821.75069 & 18.003 & 0.010 & 8.84 & 9.80(15) \\
NGC 6192 & C1636$-$432 & 64 & $-$ & 38 & 6 & 2449822.79583 & 44.949 & 0.006 & 7.95 & 11.15(20)  \\
NGC 6208 & C1645$-$537 & 15 & $-$ & 12 & 2 & 2449880.69236 & 2.010 & 0.020 & 9.00 & 10.00(15) \\
NGC 6396 & C1734$-$349 & 48 & 2 & 18 & 5 & 2449821.88958 & 31.939 & 0.014 & 7.40 & 10.60(15)  \\
NGC 6451 & C1747$-$302 & 41 & 2 & 12 & 2 & 2449883.78125 & 0.988 & 0.010 & 8.30 & 11.65(20) \\
NGC 6611 & C1816$-$138 & 45 & 2 & 42 & 5 & 2449849.79306 & 38.989 & 0.016 & 6.48 & 11.65(10) \\
NGC 6705 & C1848$-$063 & 275 & 1 & 43 & 5 & 2449822.86250 & 59.911 & 0.014 & 8.40 & 11.65(20) \\
NGC 6756 & C1906+046 & 33 & 3 & 38 & 5 & 2449823.90069 & 53.944 & 0.008 & 8.11 & 12.60(15) \\
\hline
NGC 3114 & C1001$-$598 & 181 & 7 & 50 & 4 & 2451138.82296 & 2.979 & 0.022 & 8.48 & 9.60(15)  \\
Collinder 272 & C1327$-$610 & 45 & 2 & 22 & 1 & 2452144.48472 & 0.020 & 0.008 & 7.11 & 11.85(15)  
\\
Pismis 20 & C1511$-$588 & 178 & 2 & 80 & 2 & 2452143.52872 & 1.045 & 0.022 & 6.70 & 12.55(20)  \\
NGC 6204 & C1642$-$469 & 268 & 3 & 55 & 1 & 2452143.63456 & 0.067 & 0.020 & 8.30 & 10.40(25)  \\
Lyng\aa\, 14 & C1651$-$452 & 60 & 3 & 70 & 1 & 2452144.58135 & 0.087 & 0.008 & 6.00 & 12.05(15)  
\\
\hline
\multicolumn{11}{l}{N$_{S}$ $\dotfill$ number of investigated stars; N$_{V}$ $\dotfill$ number of
variable objects; N$_{F}$ $\dotfill$ number of frames;} \\
\multicolumn{7}{l}{N$_{N}$ $\dotfill$ number of nights; $\Delta t$ $\dotfill$ time base
of the observations} \\
\end{tabular}
\end{center}
\end{table*}

\section{Observations and reductions}


The observations of the open clusters were performed 
with the Bochum 61\,cm (ESO-La Silla), the Helen-Sawyer-Hogg 
61\,cm (UTSO-Las Campanas Observatory) and  
the Complejo Astron\'omico el Leoncito (CASLEO) 2.15\,m telescopes.
The characteristics of the instruments can be found in
Bayer et al. (2000) and Paunzen et al. (2002).

The basic reductions (bias-subtraction, dark-correction, 
flat-fielding) were carried out within standard IRAF routines.
For all frames we applied a point-spread-function (PSF) fitting within the
IRAF task DAOPHOT (Stetson 1987). 

All observations were done in the intermediate
band, three filter $\Delta a$ system.
It consists of the filters $g_{\rm 1}$ ($\lambda_C$ = 5000~{\AA}, 
bandwidth = 130~{\AA}), $g_{\rm 2}$ (5220~{\AA}, 130~{\AA}) and $y$ (5500~{\AA},
230~{\AA}). The filter transmission curves and more details about  
the actual photometric
system can be found in Maitzen et al. (1997) and Kupka et al. (2003).

\begin{figure*}
\begin{center}
\includegraphics[scale=0.63]{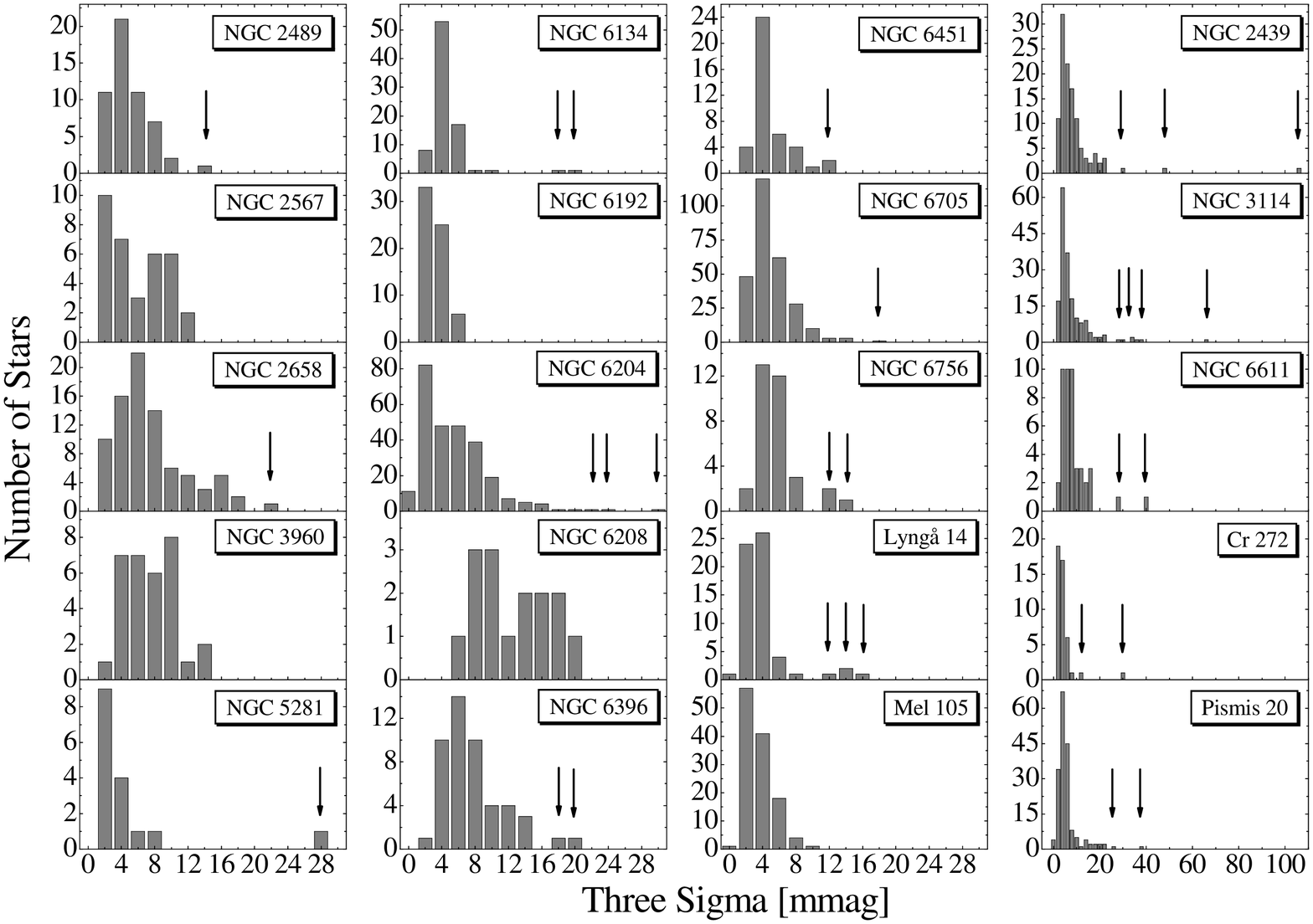}
\caption{Histograms of the three sigma standard
deviations of all mean photometric values for the programme clusters;
arrows indicate the bona-fide variable stars detected.}
\label{histos}
\end{center}
\end{figure*}

\begin{table}
\caption[]{Photometric data for the Johnson $UBV$ system taken from WEBDA for the 
identified variable stars; the mean reddening values for the open clusters are according to
the literature.
The lower panel includes the four variable objects which are most probably not
members of the corresponding cluster.
The objects are numbered according to WEBDA or to our internal numbering
system (marked with asterisks).} 
\label{variables}
\begin{center}
\begin{tabular}{llcrrr}
\hline
\hline
		 \multicolumn{1}{c}{Name} & \multicolumn{1}{c}{No.} & $E(B-V)$ & \multicolumn{1}{c}{$V$} & 
$B-V$ & \multicolumn{1}{c}{$U-B$} \\
\hline
NGC 2439 & 128 & 0.37 & 14.72 & 0.33 & 0.22 \\
         & 239 &      & 14.84 & 0.63 & 0.41 \\
         & 741 &      & 15.91 & 1.61 \\
NGC 2489 & 28 & 0.40 & 15.46 & 0.93 & 0.21 \\
NGC 2658 & 56$^{\ast}$ & 0.44 & 16.73 & 0.64 \\
NGC 3114 & 14$^{\ast}$  & 0.07 & 15.51 & 0.75 \\
         & 143$^{\ast}$ &      & 15.83 & 0.64 \\
		 & 193$^{\ast}$ &      & 15.70 & 0.74 \\
         & 233$^{\ast}$ &      & 16.29 & 0.70 \\
	     & 272$^{\ast}$ &      & 16.17 & 0.49 \\
         & 273$^{\ast}$ &      & 15.86 & 0.96 \\
         & 274$^{\ast}$ &      & 10.19 & 0.13 \\ 
Cr 272 & 1287 & 0.45 & 18.43 & 1.27 \\
       & 1297 &      & 14.21 & 0.57 \\
NGC 5281 & 1435 & 0.26 & 13.47 & 0.40 \\
Pismis 20 & 28 & 1.25 & 16.04 & 1.17 \\
NGC 6134 & 42 & 0.36 & 12.40 & 0.56 & 0.49 \\
         & 662 &     & 15.14 & 0.50 &      \\
NGC 6204 & 92$^{\ast}$  & 0.45 & 16.17 & 1.49 \\
         & 150$^{\ast}$ &      & 14.80 & 1.10 \\
         & 360$^{\ast}$ &      & 14.99 & 1.02 \\
Lyng\aa\,14 & 115 & 1.48 & 14.99 & 1.38 \\
            & 150 &      & 14.78 & 1.21 \\
NGC 6396 & 1  & 0.97 & 9.79  & 1.82 & $-$0.06 \\
NGC 6451 & 199 & 0.70 & 14.06 & 0.62 \\
         & 716 &      & 14.46 & 0.71 &  \\
NGC 6611 & 198 & 0.85 & 13.21 & 0.60 & $-$0.06 \\
         & 343 & 1.16 & 11.72 & 0.87 & $-$0.17 \\
NGC 6705 & 770 & 0.43 & 13.72 & 0.50 & 0.42 \\
NGC 6756 & 21 & 0.70 & 14.44 & 1.11 & \\
		 & 24 &      & 14.50 & 1.11 & \\
\hline
Pismis 20 & 27 & $\approx$0 & 15.43 & 0.03 \\
Lyng\aa\,14 & 101 & $\approx$0 & 14.39 & 0.62 \\
NGC 6396 & 20 & high & 10.62 & 2.44 & 2.63 \\
NGC 6756 & 40 & $\approx$0 & 15.01 & $-$0.01 & \\
\hline
\end{tabular}
\end{center}
\end{table}

\section{Temporal analysis and results} \label{time}

The temporal analysis of our photometric data is especially sophisticated
since the overall time bases range from 0.020 to 60 days with
12 to 80 data points per individual cluster (Table \ref{clusters}).
The smallest time resolution
is about one minute with typically six frames within 30 minutes.
As a consequence of the aspects discussed in the following, we do not
present any light curves. 

A classical time series analysis such as a Fourier technique
(Handler et al. 2003) cannot be performed since it is not optimized for
sparse data sets with widely different time bases. 
We have used the following
approach to get a statistically solid limit for variability.

Since we only have a limited amount of available data, all observations
were added and analysed together.
For each frame we get a ``standard mean magnitude'' as the weighted mean photometric value 
(the weights are the measurements errors according to the performed
PSF reduction technique) of all objects (variable 
and non-variable). 
The mean atmospheric extinction within the corresponding 700\,\AA\, 
decreases slightly with $\lambda$ and may vary during the time scale
of our observations (Schuster \& Parrao 2001). Furthermore, the quantum efficiency of CCD 
detectors
increases towards the red region. We are therefore confronted with different zero points
for the different standard mean magnitudes. The light curve of the ``overall standard
star'' was used as comparison in the further analysis.

As final step, differential light curves of each individual object in comparison to
the ``overall standard star'' were generated. For all differential light curves, a mean
magnitude and its standard deviation were calculated.
We define an object as variable if
\begin{itemize}
\item its standard deviation from the mean exceeds nine times the overall standard deviation of 
the cluster
\item at least three data points exceed three times its standard deviation.
\end{itemize}
The first term is only the formulation of the statistical significance whereas the
second one guarantees that bad measurements or possible misidentifications do not
affect our conclusions. The values of the mean values are between
0.006 and 0.022\,mag (Table \ref{clusters}). As a test, a phase dispersion minimization
(Stellingwerf 1978) analysis was performed yielding the same statistical significance
of variability. But the very unfavourable spectral window of the data sets prevents a decisive
conclusion about the true periods.

Figure \ref{histos} shows the histograms of the three sigma standard
deviations of all mean photometric values of the participating cluster
stars. The plotted standard deviations were normalized according
to the errors of the photon noise. 
The influence of the number of data points on the detection limit
is clearly visible (e.g. NGC 5281). We find no correlation between the amount
of detected variable objects and the value of the detection limit. 

\begin{figure}
\begin{center}
\includegraphics[width=0.4\textwidth, angle=270]{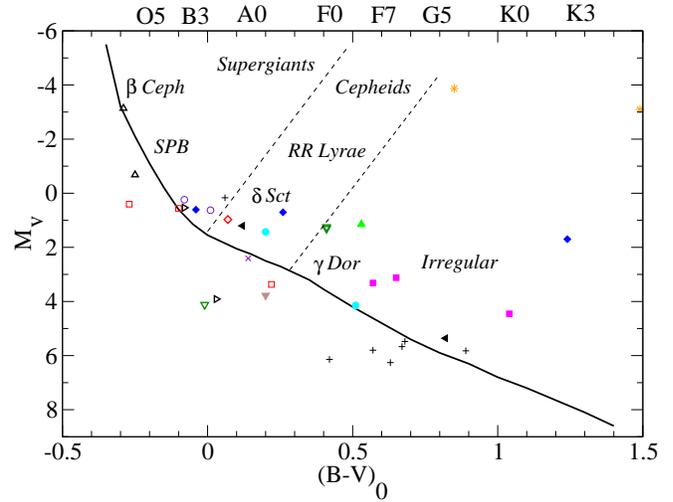}
\caption{The Hertzsprung-Russell-diagram of the variable objects with the
parameters listed in Table \ref{variables}. The zero age main sequence ist taken
from Schmidt-Kaler (1982). The areas of different variable groups are indicated.}
\label{hrd}
\end{center}
\end{figure}

\subsection{Individual variables}

Table \ref{variables} lists the 35 bona-fide variable objects found in
15 open clusters. The photometric data for the Johnson $UBV$ system were
taken from WEBDA (accessible via http://obswww.unige.ch/webda/). 
Figure \ref{hrd} shows the Hertzsprung-Russell-diagram of these objects with the
parameters listed in Table \ref{variables}. The zero age main sequence (ZAMS hereafter)
is taken from Schmidt-Kaler (1982). We are able to conclude from Table \ref{variables}
and Fig. \ref{hrd} that the following objects are {\it not} members
of the according open clusters because with the given $B-V$ color and reddening, 
they would lie significantly below the ZAMS (Fig. \ref{hrd}):
Pismis 20\,\#\,27, Lyng\aa\,14\,\#\,101 and NGC 6756\,\#\,40. However, with a reddening
close to zero, they are very close to the ZAMS indicating that these objects are foreground
stars. NGC 6396\,\#\,20 seems to be a highly reddened background star.
All other objects are probably members of the corresponding clusters; although no
further membership criteria were found in the literature. We have to emphasize that
the errors of $B-V$ are in the order of 0.05 to 0.3\,mag depending on the
type of measurement technique (photographic, photoelectric or CCD).

We have searched the literature if variable objects have been
published for the investigated clusters in the past. Only NGC 6134
was so far investigated in this respect (Rasmussen et al. 2002). They
included also the data used in this survey to show that the identified variable
stars listed in Table \ref{variables} are indeed known $\delta$ Scuti type
objects with the possibility of $\gamma$ Doradus nature. Since the detection
limit of NGC 6134 is defined as 0.010\,mag, we are confident that our method
is valid for the given data sets.  

None of the objects with a peculiar $\Delta a$ value shows a variable
nature.

Another important point is the nature of variability. In Fig. \ref{hrd} we
have indicated the position of known variable star groups. Most of the objects seem to
lie within the classical instability strip. Another large group is located
within the area of the irregular variables (e.g. T Tauri objects). However,
for an unambiguous conclusion about the true nature of the detected variability,
follow-up observations are needed.

\begin{acknowledgements}
We acknowledge partial support by the Fonds zur F\"orderung der
wissenschaftlichen Forschung, project P14984.
The CCD and data acquisition system at CASLEO has been partly financed
by R.M. Rich through U.S. NSF Grant AST-90-15827.
Use was made of the SIMBAD database, operated at CDS, Strasbourg, France and
the WEBDA database, operated at the Institute of Astronomy of the University
of Lausanne. This work benefitted from the financial contributions of 
the City of Vienna
(Hochschuljubil\"aumsstiftung projects: Wiener Zweikanalphotometer and
H-112/95 Image Processing).
\end{acknowledgements}

\end{document}